\documentclass[aps,onecolumn,showpacs,preprintnumbers]{revtex4-2}
\usepackage{float}
\usepackage{dblfloatfix}
\usepackage{fixltx2e}
\restylefloat{table}
\usepackage{graphicx}  
\usepackage{subfig}
\usepackage{subfloat}
\usepackage{multirow}
\usepackage{fancyhdr}
\usepackage{parskip}
\usepackage[T1]{fontenc}
\usepackage{dcolumn}   
\usepackage{bm}        
\usepackage{amsfonts}  
\usepackage{amsmath}   
\usepackage{amssymb}   
\usepackage{makecell}
\usepackage{caption}
\usepackage{xcolor}
\DeclareMathOperator\erf{erf}
\newcommand{\Lim}{\displaystyle\lim}

\begin{document}

\title{Physical interpretation of It\^o--distribution on the basis of local measurement of diffusion}

\author{A. Bhattacharyay}
\email{a.bhattacharyay@iiserpune.ac.in}

\affiliation {\it Indian Institute of Science Education and Research, Pune, India}

\date{\today}

\begin{abstract}  
In this paper we provide a physical interpretation of It\^o-process resulting in thermal equilibrium distribution of a Brownian particle experiencing coordinate dependent diffusion. Since the local quantities like diffusivity would go through large fluctuations in thermal equilibrium, one needs to take these fluctuation into account. We identify that the definition of local diffusivity is an essential ingredient that effectively modifies Hamiltonian of the system to result in a physically relevant Gibbs measure related to the It\^o-distribution.  

\end{abstract}


\maketitle 
\section{Introduction}
The problem of equilibrium of a Brownian particle with coordinate dependent diffusion is not yet resolved. However, consideration of coordinate dependence of diffusion is a reality which is felt over a wide spectrum of mesoscopic systems \cite{roussel2004reaction,barik2005quantum,sargsyan2007coordinate,chahine2007configuration,best2010coordinate,lai2014exploring,berezhkovskii2017communication,foster2018probing,ghysels2017position, yamilov2014position}. The controversy is primarily around the form of correct equilibrium distribution of a Brownian particle (BP) undergoing coordinate dependent diffusion -- whether it is an It\^o-distribution of the form $\frac{N}{ D(x)}e^{-U(x)/k_BT}$ or the Boltzmann distribution of the form $Ne^{-U(x)/k_BT}$ in a confining potential $U(x)$ at temperature $T$ with the Boltzmann constant being $k_B$ and the normalization factor $N$. 

Often this problem is viewed as an issue of choosing It\^o or Stratonovich convention to work out the non-linear stochastic equations where physicists prefer the latter convention while mathematicians go by the former. It\^o-convention implements the process of Brownian motion in its all rigour to the stochastic dynamics. So, it is practically unquestionable that the result of It\^o-process would come out to be consistent with Brownian motion under coordinate-dependent diffusion. Stratonovich-convention is in general employed to physics problems involving coordinate-dependent diffusion under conditions which are non-equivalent to an It\^o-process to get a Boltzmann distribution for equilibrium with a heat bath \cite{lau2007state,sancho1982adiabatic,sancho2011brownian,farago2014langevin,farago2014fluctuation}. The goal achieved in such exercises is getting the Boltzmann distribution for a confined BP at the cost of introducing correlations in the thermal noise.

There have been many attempts to get the understanding of physical situations corresponding to various conventions for non-linear stochastic processes. In particular, a generalization of Stratonovich-convention which is known as H\"anggi-Klimontovich--convention also produces Boltzmann distribution for a BP with coordinate dependent diffusion. Hänggi–Klimontovich has been compared to It\^o-process by Sokolov considering slow modulation of periodic potential \cite{sokolov2010ito}. Itô, Stratonovich and Hänggi–Klimontovich conventions have been discussed in relation to an infinite density (non-normalizable state) and its shape by Leibovich and Barkai in \cite{leibovich2019infinite}. It is argued by Tupper and Yang that specification of coordinate dependence of diffusivity needs to be accompanied by additional information to specify the state \cite{tupper2012paradox}.

Recently, Bhattacharyay and co-workers have argued in favour of It\^o-distribution as opposed to the Boltzmann-distribution for thermal equilibrium a BP undergoing coordinate-dependent diffusion in confinement \cite{bhattacharyay2019equilibrium, bhattacharyay2020generalization, maniar2021random}. There has been an attempt made by bhattacharyay \cite{bhattacharyay2020generalization} to show that the It\^o-distribution could be consistent with Gibbs measure $P(x,v) \sim e^{-\left(\frac{mv^2}{2}+U(x)\right)/k_BT}$ if one uses a velocity upper limit, while integrating out the factor $e^{-\frac{mv^2}{2k_BT}}$, proportional to local diffusivity $D(x)$. In the present paper, we extend this approach to make the It\^o-distribution be consistent with local validity of equipartition. We identify that the local inverse diffusivity considered in It\^o-processes is actually a fluctuation averaged quantity $\langle 1/D(x)\rangle$. We demonstrate restrictions imposed by observation of local diffusivity in the theory of such systems. We show that, with these essential ingredients taken into account, the It\^o-process finds physical interpretation and its distribution is consistent with physically meaningful Gibbs measure. The method we present here is very general. Fluctuating (or diffusing) diffusion is an idea of contemporary interest \cite{chubynsky2014diffusing,cherstvy2016anomalous,chechkin2017brownian,paul2018reaction,jain2017diffusing} and the present work could be of relevance to the community working on such problems as well.

Diffusivity (or diffusion constant) in the uniform case gets defined to be proportional to the average rate of mean of the squared displacement at the infinite time ($t\to\infty$) limit. This definition cannot hold in its totality when the diffusivity is a local variable because one cannot implement the $t\to\infty$ limit in which the particle moves around everywhere. Average time-scale over which the local diffusivity gets defined would depend on the length scale of the locality and obviously there would be fluctuations around this average. In the local case, therefore, the $t\to\infty$ limit will get replaced by an ensemble average on fluctuations. Even in the presence of coordinate dependent diffusion, the velocity distribution is Maxwellian everywhere in space. The time integral of velocity autocorrelation produces diffusivity. Therefore, diffusivity varies over places where the distribution of velocity is always Maxwellian. The only place where the information of local diffusivity can enter the velocity distribution, then, is its upper cut-off. In any realistic case, there must be a finite upper cut-off velocity to the distribution. It could be neglected in favour of an arbitrarily high number in the uniform diffusion case because, it would get absorbed in the normalization constant in the end. However, it is not the case when diffusivity is coordinate dependent. We show that, identification of the local velocity cut-off in relation to the local diffusivity and ensemble average over fluctuations will reveal the Hamiltonian corresponding to the Gibbs measure for the It\^o-distribution. 

The Hamiltonian of BP has to modify in the presence of coordinate dependence of diffusion because that coordinate dependence of diffusion results from other degrees of freedom which are hydrodynamic in nature. This is why a BP shows coordinate dependence of diffusion near a wall \cite{faucheux1994confined}. Local influence of those hydrodynamic degrees of freedom gets effectively included in the stochastic dynamics of the BP through a local modification of the diffusion. This, in every likelihood, should result in the presence of an additional diffusivity dependent interaction term in the Hamiltonian of the BP to take into account hydrodynamic degrees of freedom not explicitly considered. A physically meaningful Gibbs measure for such a system could not be found out until we identify this additional interaction term in the Hamiltonian of the BP.

\section{The Hamiltonian and the distribution}
To get correspondence between It\^o-distribution and physically relevant Gibbs measure, we need to identify the appropriate Hamiltonian. We are at liberty to choose a diffusivity-dependent effective potential for this purpose if that could be justified on physical ground. We would identify in the following a Hamiltonian of the form
\begin{equation}
H=\frac{p^2}{2m} + (U(x) + F[D(x)])
\end{equation}
of a BP seeing an interaction potential $(U(x)+F[D(x)])$ in one dimensional space where its diffusivity $D(x)$ is coordinate dependent. The BP of mass $m$ has velocity $v$ with $p=mv$. $F[D(x)]$ is the diffusivity dependent part of Hamiltonian which is to be found out. With the knowledge of interaction potential $U(x)$, one conventionally moves from $e^{-H/k_BT}\to e^{-U(x)/k_BT}$ in the overdamped limit by taking $m\to 0$. We will see in what follows that, the It\^o-distribution $N\left\langle\frac{1}{D(x)}\right\rangle e^{-U(x)/k_BT}$ ($N$ is normalization constant) results at the overdamped limit of a BP with coordinate dependent diffusion where $\langle 1/D(x)\rangle$ is fluctuation averaged inverse of local diffusivity. 

Consideration of a local distribution of the diffusivity makes sense because it is a local thermodynamic quantity and must undergo fluctuations. One can expect the structure of function $F[D(x)]$ is such that it should vanish for (1) uniform diffusion (diffusivity is a constant), (2) it must vanish at the overdamped limit $m\to 0$ and (3) it must also vanish for $T=0$. We will see that all these demands on $F[(D(x))]$ can be met.

The velocity distribution factor $e^{-\frac{mv^2}{2k_BT}}$ is clearly flat with the value unity at the limit $m\to 0$. Therefore, all the velocities are equally likely at this limit. However, this is an un-physical distribution without a physically motivated upper cut-off velocity. This requires setting an upper cut-off for local normalization of velocity distribution such that it remains related to the local diffusivity. Question is, how to set this upper cut-off of the velocity based on local information. In order to set this, the actual experimental protocol, keeping the presence of thermal fluctuations in mind, can help.

Let us think about the possible protocol of defining local diffusivity. If one tries to locally measure the diffusivity, one has to find how long ($\delta t_i(x)$) it takes for a BP to reach a small distance $l$ (say) starting form the position $x$ at time $t_i$. The length scale $l$ must be small and the same everywhere over space so that local diffusivity remains defined on the same footing everywhere in space. This is the discretization of space at the same scale everywhere in order to be able define coordinate dependence of diffusivity and damping.

This way, the $i$th measurement gives the diffusivity to be $D_i(x)=\frac{l^2}{2\delta t_i(x)}$ at an average velocity $u_i(x)=\frac{l}{\delta t_i(x)}=\frac{2D_i(x)}{l}$ over the interval $\delta t_i(x)$. Note that, $D_i(x)$ and $\delta t_i(x)$ are dependent stochastic variables by the definition of $D_i(x)$. The interval $\delta t_i(x)$ is obviously a fluctuating quantity which will set fluctuations in the $u_i(x)$ and $D_i(x)$. The local measure of diffusivity must be an ensemble average $\langle D(x)\rangle=\lim_{n\to \infty}\sum_{i=1}^n{D_i(x)/n}$. Diffusivity has to be defined locally in this way because we cannot take $t\to\infty$ limit in defining local diffusivity as is usual for uniform diffusion. The ensemble average at $n\to\infty$ plays the role of the $t\to\infty$ of the uniform diffusion. The underlying assumption to this ensemble average is the fact that the particle visits position $x$ many times during its excursion.

The same applies to the damping $\langle\Gamma(x)\rangle=\lim_{n\to \infty}\sum_{i=1}^n{\Gamma_i(x)/n}$. Diffusivity and damping being related by the Stokes Einstein relation in each interval of time $\delta t_i(x)$, $D_i(x)\Gamma_i(x)=k_BT$, one can expect the ensemble average to relate these as $\langle\Gamma(x)\rangle = k_BT\left\langle \frac{1}{D_i(x)}\right\rangle $. Now, knowing the $\langle \Gamma(x)\rangle$ from the hydrodynamic theory one would also know the $\left\langle\frac{1}{D(x)}\right\rangle = \frac{\langle\Gamma(x)\rangle}{k_BT}$. We have to use the velocity measured over a time $\delta t_i(x)$
\begin{equation}
u_i(x)=\frac{2D_i(x)}{l},
\end{equation}
as the cut-off for integration of the velocity distribution keeping in mind that we need to do an ensemble average over this cut-off. Actually, this is the average velocity (with its fluctuations) that the BP registers on the discretized space of length scale $l$ in each interval $\delta t_i(x)$.

One could have also taken any finite constant multiple of $u_i(x)$ to be the upper cut-off, however, on ensemble average the constant factor would not have any effect. Length scale $l$ could be chosen related to the other physics (may be hydrodynamics) underlying coordinate dependence of diffusion. However, the cut-off being proportional to $u_i(x)$ is crucial because that is what the system gets to see over the length scale $l$. Identifying this $l$ exactly in the context of overdamped limit is not required because this constant will ultimately be absorbed in the overall normalization factor. 

Let us have a look at the local equipartition of kinetic energy with the normalized velocity distribution at the overdamped limit. Consider
\begin{equation}
	\int_{-u(x)}^{u(x)}{dv e^{-mv^2/2k_BT}}=\sqrt{\frac{2k_BT}{m}}\int_{-u(x)\sqrt{\frac{m}{2k_BT}}}^{u(x)\sqrt{\frac{m}{2k_BT}}}{dze^{-z^2}}=\sqrt{\frac{2\pi k_BT}{m}}\erf{\left (u(x)\sqrt{\frac{m}{2k_BT}} \right)}=N(x)
\end{equation}
where we have got rid of the index $i$ in the notation of velocity. Thus, the average squared-velocity due to the Maxwellian distribution within interval between cut-off $\pm u(x)$ is

\begin{eqnarray}\nonumber
\langle v^2\rangle_{u(x)} &=& \Lim_{m\to 0} \frac{2}{N(x)}\int_0^{u(x)}{v^2e^{-mv^2/2k_BT}dv}\\\nonumber
&=& \Lim_{m\to 0} \frac{1}{N(x)}\left [ \frac{\sqrt{\pi}\erf\left(u(x)\sqrt{\frac{m}{2k_BT}}\right)}{2\left (\frac{m}{2k_BT}\right )^{3/2}} - \frac{u(x)e^{-\frac{mu(x)^2}{2k_BT}}}{\frac{m}{2k_BT}}\right]
\\ &=& \frac{1}{2u(x)}\left [ \frac{u(x)}{\frac{m}{2k_BT}} -\frac{u(x)\left (1-\frac{mu(x)^2}{2k_BT}\right )}{\frac{m}{2k_BT}}\right ] = \frac{u(x)^2}{2},
\end{eqnarray}
where $N(x)=2u(x)$ in the overdamped limit. Now, the result shown in equation (4) cannot obviously satisfy equipartition because of the coordinate dependence of $u(x)=2D(x)/l$ and this is an average within a particular time interval $\delta t(x)$. However, we have already considered $D(x)$ to have fluctuation whose density should be of the form $e^{-F[D(x)]/k_BT}$. An ensemble average is due at this stage over these fluctuation. If we keep in mind that $u(x)$ is itself an average quantity over the interval of time $\delta t(x)$ which has resulted from a sum over many independent velocities on even smaller time intervals within $\delta t(x)$, then, the central limit theorem applies on $u(x)$. Therefore, take the reasonable consideration that, fluctuations result in a normal distribution of $u(x)$ at every $x$ of the form
\begin{equation}
P[u(x)]= \frac{1}{\sqrt{2\pi}\sigma}e^{-\frac{1}{2}\left(\frac{u(x)-\langle u(x)\rangle}{\sigma} \right )^2},
\end{equation}
where $\langle u(x)\rangle$ is also a function of $x$. Doing an average of $\langle v^2\rangle_{u(x)}$ over this distribution we finally get
\begin{equation}
\langle v^2\rangle_{u(x)} = \frac{\langle u(x)^2\rangle}{2}= \frac{1}{2}(\sigma^2 + \langle u(x)\rangle^2).
\end{equation}

The equipartition now demands
\begin{equation}
\sigma^2=\frac{2k_BT}{m}-\langle u(x)\rangle^2,
\end{equation}
at the overdamped limit, the r.h.s of equation (7) is safely a positive number considering $\langle u(x)\rangle$ finite everywhere in space. In view of equation (5), the local normalized distribution of the fluctuations of $D(x)$ can now be written as
\begin{eqnarray}\nonumber
P\left [\frac{D(x)}{l}\right ] 
&=& \frac{1}{\sqrt{2\pi}\sigma}e^{-\frac{1}{2}\left(\frac{u(x)-\langle u(x)\rangle}{\sigma} \right )^2}\\&=&\frac{1}{\sqrt{4\pi\left(\frac{k_BT}{m}-2\left\langle\frac{D(x)}{l} \right\rangle^2 \right)}}\left [e^{-\frac{\left [\frac{D(x)}{l}-{\left\langle\frac{D(x)}{l}\right \rangle}\right ]^2 }{\frac{k_BT}{m}-2\left\langle\frac{D(x)}{l}\right \rangle^2}}\right].
\end{eqnarray}

This, readily makes us identify
\begin{equation}
F[D(x)]={\frac{mk_BT\left [\frac{D(x)}{l}-{\left\langle\frac{D(x)}{l} \right\rangle}\right ]^2 }{k_BT-2m \left\langle \frac{D(x)}{l}\right \rangle^2}}.
\end{equation}
We can identify three important limits -- at (1) $T \to 0$, (2) $m \to 0$ and (3) $D(x)\to$ constant, $F[D(x)]\to 0$ such that the Hamiltonian retains its form $H = p^2/2m + U(x)$. We would look at the implication on dynamics in the following after seeing the emergence of the It\^o-distribution from this measure.

The full distribution up to a normalization factor now becomes
\begin{equation}
P(x,u(x),v) \sim e^{-U(x)/k_BT}\times P[u(x) ]\times \frac{1}{N(x)}e^{-mv^2/2k_BT}
\end{equation}

Taking the limit $m \to 0$ where $e^{-mv^2/2k_BT}\to 1$ and integrating out $u(x)$, we get the distribution as

\begin{equation}
P(x)\sim e^{-U(x)/k_BT}\times \int_{-\infty}^\infty{d[u(x)]P[u(x)]\frac{1}{2u(x)}}=\frac{1}{2}\left\langle\frac{1}{u(x)}\right\rangle e^{-U(x)/k_BT}
\end{equation}
Now, identifying $u(x)=2D(x)/l$, we finally get
\begin{equation}
P(x)=N\left\langle\frac{1}{ D(x)}\right\rangle e^{-U(x)/k_BT},
\end{equation}
where the distribution is now normalized with the normalization constant $N$ over all space. 
This we can easily identify with the It\^o-distribution by acknowledging the fact that, the local coordinate dependent diffusivity that this process takes into account is an ensemble average of local diffusivity measured many times over a small length scale $l$ at position $x$. We also identify the fact that fluctuations of the local diffusivity follow a normal distribution under equilibrium thermal fluctuations and the Hamiltonian has been modified accordingly to take into account those fluctuations. Had the Hamiltonian not been modified to take into account the fluctuations of local diffusivity there is no scope for those to enter the structure of dynamics which practically includes $\langle\Gamma(x)\rangle$ as a measured/estimated quantity.

The general dynamics including the inertial term, which would be consistent with the above mentioned It\^o-distribution at the overdamped limit, be written as
\begin{equation}
m\frac{dv}{dt} = -\langle m\zeta(x) \rangle v- \frac{\partial}{\partial x}(U(x)+ F[D(x)]) +\langle m\zeta(x)\rangle\sqrt{2/\left\langle\frac{1}{D(x)}\right\rangle}\eta(t),
\end{equation}
where $m\zeta(x)=\Gamma(x)$ and the overdamped limit has to be understood as $m\to 0$ and $\zeta(x)\to \infty$ such that $\Gamma(x)$ is finite. This dynamics will result in the distribution as shown in equation (10), keeping in mind the form of $F[D(x)]$ as given in equation (9). At the overdamped limit the dynamics is of the form

\begin{equation}
\frac{dx}{dt}=-\frac{1}{\langle \Gamma(x)\rangle} \frac{\partial U(x)}{\partial x} +\sqrt{2/\left\langle\frac{1}{D(x)}\right\rangle}\eta(t),
\end{equation}
which by the correct stochastic analysis (using It\^o-convention or equivalent Stratonovich-convention for equilibrium Brownian motion) will result in the distribution as shown in (12) when we keep in mind the Stokes-Einstein relation being written as $\langle\Gamma(x)\rangle = k_BT\left\langle \frac{1}{D(x)}\right\rangle$. Moreover, the local ensemble averaged Stokes-Einstein relation has followed from $\Gamma(x) D(x)=k_BT$ which has been holding within each time interval $\delta t(x)$. 

\section{Discussion}
It\^o-process for a nonlinear stochastic problem finds an adequate physical interpretation by the inclusion of local diffusivity on the velocity distribution cut-off. We identify the fact that, local diffusivity considered in the nonlinear stochastic problem of an It\^o-process has to be the ensemble-averaged diffusivity. This diffusivity is related to the ensemble-averaged damping coefficient by the Stokes-Einstein relation $\langle\Gamma(x)\rangle=k_BT\left\langle\frac{1}{D(x)} \right\rangle$. The It\^o-process and resulting equilibrium distribution come out to be consistent with physically meaningful Gibbs measure when the above mentioned definition of the local ensemble averaged damping and diffusivity are at work. One cannot just go by the usual definition of diffusivity as applicable to a constant diffusion because the $t\to\infty$ limit is not applicable in the local case. The Hamiltonian is also consistently modified with all the required limits on it, namely the overdamped limit, the zero temperature limit and the constant diffusivity limit. The modification is effectively taking into account the fluctuations of local diffusivity to remain bounded by a harmonic interaction limited by the temperature of the system.

It\^o vs Stratonovich controversy, to the knowledge of the present author, has not so far been adequately focussed on the possible physical picture associated to an It\^o-process. This requires one to first understand what possible physical scenario could be associated to an It\^o-process. Stratonovich-convention is mostly preferred by physicists to the It\^o-convention presumably for the scope of manipulation with the former to get the Boltzmann distribution. As Van Kampen has pointed out \cite{van1981ito} and the same being re-iterated by others \cite{mannella2022ito}, identification of the physical process as the It\^o or Stratonovich is essential to decide which convention to use in a stochastic analysis. There hardly exists any scope for controversy until the physical processes are identified. There is nothing objectionable in the stochastic analysis as per Ito-convention from the perspective of thermal equilibrium rather this convention is perfect on the stochastic side and has reasonable physical interpretation when approached from a meaningful Gibbs measure. Moreover, one should also keep in mind that the Gibbs measure is generally applicable to all scales but not the Boltzmann distribution which is typically proved at the thermodynamic limit.

\section*{Acknowledgement} I would like to acknowledge discussions with Bertrand Halperin and J. K. Bhattacharjee which has motivated this work.

\bibliography{reference.bib}

\end{document}